\documentclass[runningheads]{llncs}
\usepackage{graphicx}
\usepackage{subcaption}

\usepackage{hyperref}

\newcommand{\module}[1]{\paragraph{\normalfont \textbf{#1}}}

\usepackage[scaled=.75]{beramono}
\usepackage[T1]{fontenc}
\newcommand{\term}[1]{\texttt{#1}}
\newcommand{\aswellas}{$\wedge$}
\newcommand{\of}[1]{_\mathrm{#1}}

\graphicspath{{./fig/}}

\begin{document}
\title{Effect of E-cigarette Use and Social Network on Smoking Behavior Change: An agent-based model of E-cigarette and Cigarette Interaction}
%
%
\author{
Yang Qin\inst{1}
\and
Rojiemiahd Edjoc\inst{2}
\and 
Nathaniel D Osgood\inst{1}
}

\authorrunning{Qin, Y. et al.}
%
\institute{ 
University of Saskatchewan, Canada\\
\email{first.last@usask.ca} \\
\and 
Statistics Canada, Ottawa, Canada\\
\email{rojiemiahd.edjoc@canada.ca} 
}
%

%
\titlerunning{An agent-based model of E-cigarette and Cigarette Interaction}

\maketitle              
\begin{abstract}
Despite a general reduction in smoking in many areas of the developed world, it remains one of the biggest public health threats. As an alternative to tobacco, the use of electronic cigarettes (ECig) has been increased dramatically over the last decade. ECig use is hypothesized to impact smoking behavior through several pathways, not only as a means of quitting cigarettes and lowering risk of relapse, but also as both an alternative nicotine delivery device to cigarettes, as a visible use of nicotine that can lead to imitative behavior in the form of smoking, and as a gateway nicotine delivery technology that can build high levels of nicotine tolerance and pave the way for initiation of smoking. Evidence regarding the effect of ECig use on smoking behavior change remains inconclusive.
To address these challenges, we built an agent-based model (ABM) of smoking and ECig use to examine the effects of ECig use on smoking behavior change. The impact of social network (\term{SN}) on the initiation of smoking and ECig use were also explored. Findings from the simulation suggest that the use of ECig generates substantially lower prevalence of current smoker (\term{PCS}), which demonstrates the potential for reducing smoking and lowering the risk of relapse. The effects of proximity-based influences within \term{SN} increases the prevalence of current ECig user (\term{PCEU}). The model also suggests the importance of improved understanding of drivers in cessation and relapse in ECig use, in light of findings that such aspects of behavior change may notably influence smoking behavior change and burden.

\keywords{E-cigarette  \and Smoking \and Agent-based modeling \and Distance-based network}
\end{abstract}
\section{Introduction}
Smoking and secondhand smoke harm nearly every organ of the body and contribute to many preventable diseases, including lung cancer, coronary heart disease, chronic obstructive pulmonary disease, and other cardiovascular diseases \cite{CenterforDiseaseCOntrolandPrevention2010,USDepartmentofHealthandHumanServices2014}. Nicotine products come in various forms, e.g., cigarettes, nicotine gum, patch, and ECig \cite{Chaturvedi_2015}. 
ECigs, vaporizing a liquid mixture which is used as a substitute for tobacco leaves and stored inside cartridges \cite{PISINGER2014248,Rom_2014}, were introduced to the market in 2003, promoted and marketed by major tobacco companies in the last decade \cite{PISINGER2014248,high_tech_approach}. 
The use of ECig as a cigarette alternative has increased dramatically. The \term{PCEU} among US adults increased from 0.3\% in 2010 to 6.8\% in 2013 \cite{Cherng_2016}. Within recent years, there has been a particularly dramatic and alarming rise in the use of ECig amongst youth.

The health behaviors associated with smoking have been studied in detail. The majority of smokers attempt to quit smoking, but fewer than 5\% of them remain quit for more than three months \cite{Axtell:2006aa}. Effective tools for smoking cessation (\term{SC}) may help current smoker (\term{CS}) quit, and forestall an individual at risk of smoking, e.g., former smoker (\term{FS}), struggling with avoiding relapse. ECigs also allow never smoker (\term{NS}) seeking to experiment with nicotine as an alternative to cigarettes. The rise of ECig use is associated with a perception that ECig is safer than cigarettes and a useful \term{SC} device. However, there remains little solid scientific evidence confirming the effectiveness and safeness of ECig as a \term{SC} tool \cite{Cherng_2016,Zhuang_2016}. By surveying 2028 US smokers in 2012 and 2014 and two-years of follow-up, Zhuang et al. \cite{Zhuang_2016} concluded that long-term ECig users had a higher rate of \term{SC} of 42.4\% than short-term Ecig users and non-users (14.2\% and 15.6\%, respectively). 
Zhu et al. \cite{Zhu:2017aa} concluded that ECig users have a higher rate of \term{SC}, and are more likely to remain quit 
than non-ECig users. 
Cherng et al. \cite{Cherng_2016} proposed an ABM to exmaine the effect of ECig on the smoking prevalence of US adults, and concluded that the simulated effects of ECig on \term{SC} largely changed smoking behavior. The ABM simulated the influences of smoking behavior on ECig use initiation and cessation, and how ECig reversely affected \term{SC} and smoking initiation \term{SI}.

While promising, previous studies have predominantly relied upon self reported surveys, cohort studies and clinical trials. Such larger studies are expensive, are associated with high delay until they show effect, and can be difficult to plan and execute given the wide variety of patterns of behavior possible (e.g., initiation of exclusive smoking following ECig use, initiation of exclusive ECig use following tobacco, dual use, start of ECig use following quitting tobacco, etc.). Clinical trials often regulate or exclude factors that play a key role in shaping outcomes in society, such as switching of nicotine delivery modality, varying rates of compliance, and peer influence effects. 

In this paper, extending the preliminary model structure introduced by Cherng et al. \cite{Cherng_2016}, we build an ABM of smoking and ECig use with modalities of initiation, cessation, and relapse to examine the effects of ECig use on individual-level smoking behavior change and population-level smoking patterns according to the aggregation of individual outcomes. Our model incorporates strong \term{SN} effects involving both selection of networks and influence over networks, age, sex and history-dependent effects regarding the rate of initiation, cessation, and relapse for both smoking and ECig use, and individual decision-making effects based on characteristics of social contacts. In particular, we use the model to investigate whether the ECig is an effective \term{SC} device and the impact of ECigs on non-smokers with regards to \term{SI}.

\section{Methods}

\module{Model Overview}
ABM can simulate complex social dynamics and behaviors with considerably high resolution, and generate population-level results by aggregating individual outcomes in different scenarios \cite{Cherng_2016}. Equally notable, ABM is widely applied to probe the impacts of counter-factual interventions, as well as to help prioritize data collection in a complex milieu of complex interactions of behaviors and product types. In this study, a high level of heterogeneity characterizing both exogenous and endogenous components, specific traits at individual level and modularity also strongly motivated the use of ABM.

Our model was built in AnyLogic (version 8.3.3), and used four interacting statecharts for each agent, featuring smoking states, ECig use states, birth, and mortality. The parameters, transition rates and statecharts in the \term{Person} class serve as influences from within an agent on smoking and ECig use behavior. The model further incorporates a distance-based network to simulate social contacts between agents. 

The model simulates a population of 100,000 agents with age distribution based on population pyramid of Canada \cite{Population_Pyramids}. The model time unit is 1yr, and the length of the time horizon is 70yrs. The initial states may misestimate the prevalence of each smoking and ECig use state, so a period of burn time (52 years) is used for the model to achieve equilibrium. Over the continuous time of the simulation, agents either maintain their current state of smoking and ECig use or transit to other status based the (hazard) rates discussed in the next section.

\module{Model Formulation}
Smoking statechart describes three smoking states: never smoker (\term{NS}), \term{CS} and \term{FS}.
An individual can switch its presence in each of the three states of statechart according to specified transition rates, namely the rate of \term{SI}, the rate of \term{SC}, and the rate of smoking relapse (\term{SR}).

ECig use statechart separates the states of ECig use as never ECig user (\term{NEU}), \term{CEU} and former ECig user (\term{FEU}). The transition of ECig use initiation (\term{ECigUI}) is fired with a hazard rate, transferring an agent from \term{NEU} to \term{CEU}.  
Other transitions are message triggered transitions, which will be activated only under scenarios when we consider: A \term{CS} who never used ECig may possibly initiate ECig use after quitting smoking, transiting from \term{CS}\aswellas\term{NEU} to \term{FS}\aswellas\term{CEU} by chance; An \term{FS} who is \term{CEU} may possibly quitting ECig after relapse to smoking, transferring from \term{FS}\aswellas\term{CEU} to \term{CS}\aswellas\term{FEU} by chance; And a \term{CS} who is \term{FEU} may possibly relapse to \term{CEU} after quitting smoking, transiting from \term{CS}\aswellas\term{FEU} to \term{FS}\aswellas\term{CEU} by chance. 
For the two statecharts, agents can occupy a specific, concrete state of one statechart at any one time, while being in any state of the other statechart. 

Rate of \term{SI}, \term{SC} and \term{SR} denoted as $r\of{si}$, $r\of{sc}$ and $r\of{sr}$, respectively, are each the product of its corresponding hazard rate 
($\alpha\of{si}$, $\alpha\of{sc}$ and $\alpha\of{sr}$ for the calculation of $r\of{si}$, $r\of{sc}$ and $r\of{sr}$, respectively), 
a multiplier 
($m\of{si}$, $m\of{sc}$ and $m\of{sr}$ for the calculation of $r\of{si}$, $r\of{sc}$ and $r\of{sr}$, respectively)
and a coefficient 
($e\of{si}$, $e\of{sc}$ and $e\of{sr}$ for the calculation of $r\of{si}$, $r\of{sc}$ and $r\of{sr}$, respectively).

The hazard rates reflect the magnitude of the effect of age, gender and smoking history on $r\of{si}$, $r\of{sc}$ and $r\of{sr}$. We transformed the annual probabilities of SI and SC ($p\of{si}$ and $p\of{sc}$, respectively) of male and female of 1970 birth cohort, reported by Holford et al. \cite{HOLFORD2014e31}, into their corresponding $\alpha\of{si}$ and $\alpha\of{sc}$ as table functions in AnyLogic by using $p = 1 - e^{-\alpha}$. The model assumed that $\alpha\of{sr}$ declines with growing time since quit; thus, individuals who only recently quit have far higher relapse risk than an agent who has remained as \term{FS} for a prolonged period. The value of multipliers is driven by the state of ECig use. Wills et al. \cite{Wills_2016} suggested that \term{NS} who tried ECig is three times more likely to start smoking. Leventhal et al. \cite{10.1001/jama.2015.8950} reported that ECig users were four times likely to uptake cigarettes. McRobbie et al. \cite{McRobbie_2014} suggested that the rate for \term{SC} was significantly higher in the presence of ECig use (RR 2.3; 95\%CI: 1.05 - 4.96). Based on the linkages between ECig use and $r\of{si}$ and $r\of{sc}$ mentioned above. As ECig use can help relieve the symptoms of nicotine withdrawal to some degree and might provide an additional avenue towards continued socialization with companions who remain tobacco users, \term{CEU}s are less likely to relapse in smoking, compared to non-ECig users \cite{Polosa_2011}. Therefore the model assumes $m\of{si}$ is 4.0 for agents who are \term{CS}\aswellas\term{CEU} \cite{10.1001/jama.2015.8950}, 
or is 2.87 for agents who are \term{FEU} \cite{Wills_2016}, 
$m\of{sc}$ is 2.3 for agents who are \term{CS}\aswellas\term{CEU} \cite{McRobbie_2014}, $m\of{sr}$ is 0.5 for agents who are \term{CS}\aswellas\term{CEU} \cite{Polosa_2011}.
If each rate is only the product of its hazard rate and a multiplier, the rate may misestimate the projection of smoking. Therefore, the coefficients $e\of{si}$, $e\of{sc}$ and $e\of{sr}$ were calibrated to match simulation outcomes against historical data.

For the rate of \term{ECigUI}, the model adapted the time-based sigmoid function and divisors 
introduced by Cherng et al. \cite{Cherng_2016}, to characterize the increasing use of ECig after its introduction into the market and the influence of smoking status on \term{ECigUI}. Additionally, the rate of \term{ECigUI} is strongly related to the agent's smoking status and demographic factors \cite{REID2015180}, suggesting that ECig is popular in smokers and young people; thus, we assumed an hazard of \term{ECigUI} of male agents using a table function, which has a x-axis of age of the agents and y-axis of the hazard rate and follows same pattern as for the hazard rate of \term{SI} for male. If an individual is female, the hazard of \term{ECigUI} of this agent is given by the corresponding point on the table function divided by the variable \term{divECigFemale} with a value of 1.5. 
The overall rate of \term{ECigUI} is the product of the hazard of \term{ECigUI} given by the time-based sigmoid function \cite{Cherng_2016} and a coefficient ($e\of{ECig}$), which was calibrated by matching model generated incidence of ECig use against corresponding historical data.

The model assumes that the transition of ECig use cessation (\term{ECigUC}) and ECig use relapse (\term{ECigUR}) are affected only by the smoking behavior, that is, the model assumed that individuals who are \term{CEU}s or \term{FEU}s would remain so unless changes occurred in their smoking behavior. Specifically, in the absence of identified evidence with respect to the fraction of individuals whose state of ECig use will be affected by smoking behavior, the model posited that 85\% of \term{CEU}s\aswellas\term{FS}s will quit ECig if they relapse to smoking, since their nicotine cravings were satisfied by smoking, and 80\% of agents who are \term{FEU}\aswellas\term{CS} will transit to \term{CEU} if they quit smoking. 
As ECigs may be used as cessation tools, the model further assumed that 50\% of smoking quitters would uptake ECig immediately after quitting smoking. Therefore, message dichotomous branching transitions were built for \term{ECigUI} and \term{ECigUC} under these assumptions in addition to the rate of \term{ECigUI} discussed above.

Age-specific birth and mortality rates drawn from Statistics Canada of 2016 \cite{birth_rate,mortality_rate} are used in the model. The total fertility rate of Canada in 2016 is 1.54 per woman. To maintain population replacement (with a total fertility rate of 2.1) for successive years of the model running, we thus multiplied a coefficient (with a value of 1.357) by the fertility rate of each age group.
\begin{figure}

\centering
\includegraphics[scale=0.28]{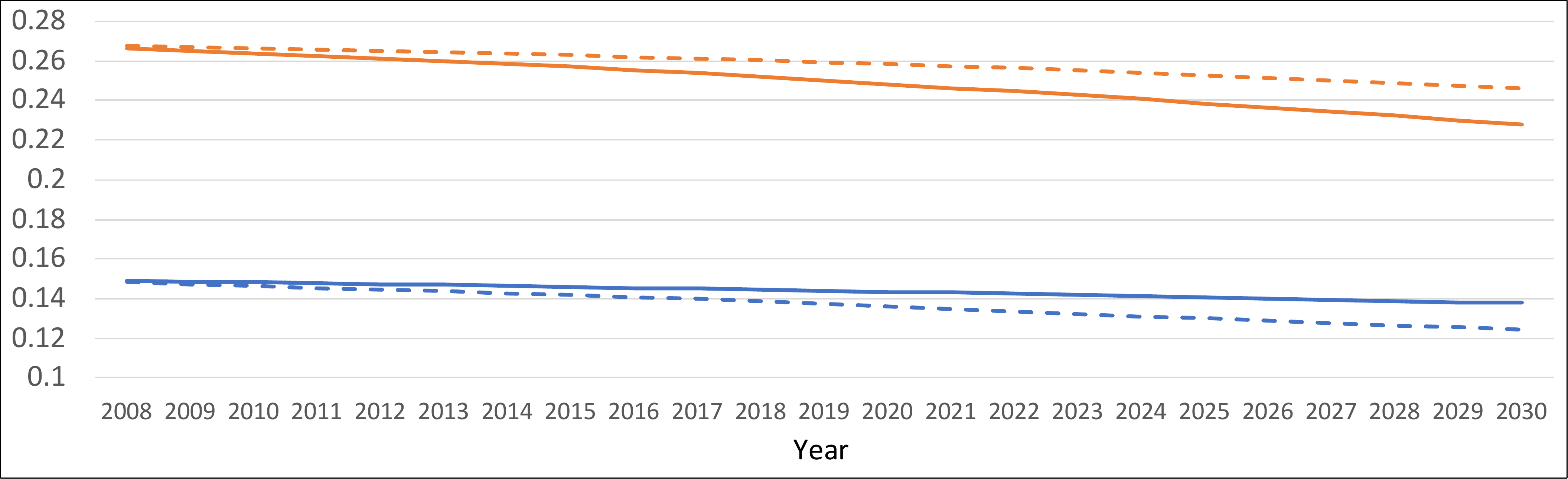}
\captionof{figure}{\term{PCS} and \term{PFS} of \term{Scn1} and \term{Scn2}. Orange and blue solid line represent the \term{PFS} and \term{PCS} in \term{Scn1}, respectively. Orange and blue dashed line represents the \term{PFS} and \term{PCS} in \term{Scn2}, respectively.}
\label{fig4}

\centering
\includegraphics[scale=0.28]{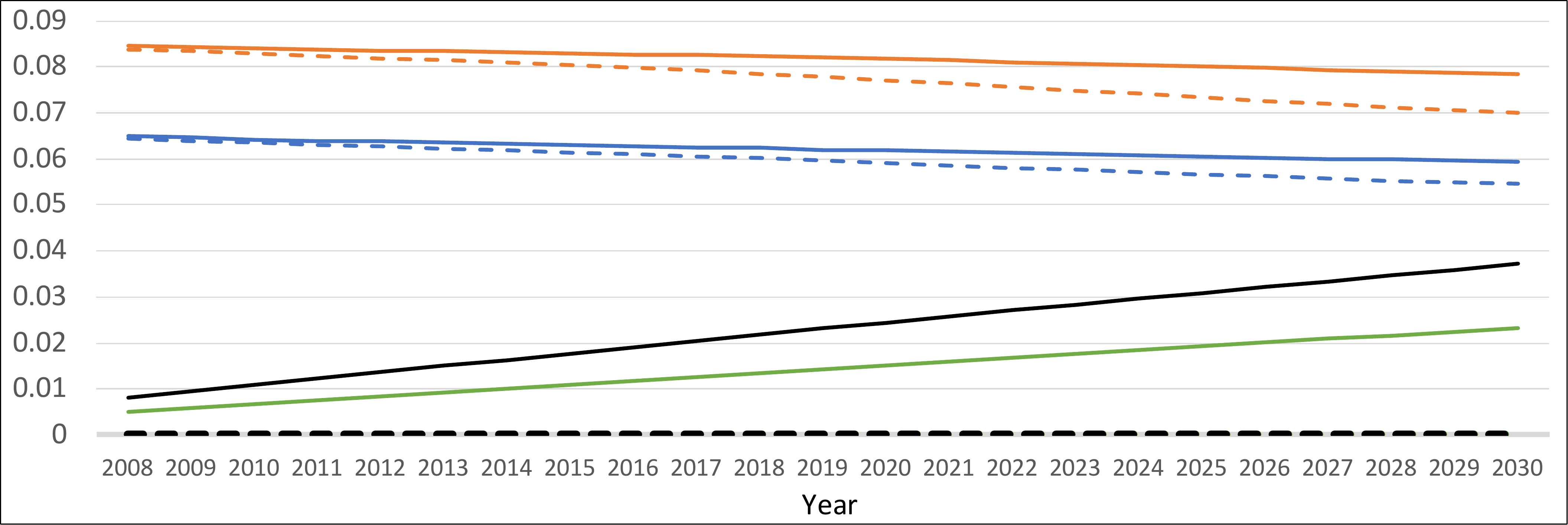}
\captionof{figure}{ \term{PFCS}, \term{PMCS}, \term{PFCEU} and \term{PMCEU} of \term{Scn1} and \term{Scn2}. Blue, orange, green, and black solid line represent the \term{PFCS} and \term{PMCS} of \term{Scn1}, \term{PFCEU} and \term{PMCEU} of \term{Scn2}, respectively. Blue, orange, green, and black dashed line represent the \term{PFCS} and \term{PMCS} of \term{Scn2}, \term{PFCEU} and \term{PMCEU} of \term{Scn1}.}
\label{fig3}

\end{figure}

Smoking is well recognized as both an individual habit and a social phenomenon \cite{Axtell:2006aa}. The baseline model was extended with a distance based network to simulate the effect of social connection and peer pressure on the \term{SI} and \term{ECigUI}.
To build a localized \term{SN} for each agent, connecting with its nearby agents, the model assumed that an agent establishes the network with the agents in proximity (50m). The \term{SN} was implemented as a dynamic network driven by agent mobility in continuous space with width and height both equal to 250,000m. Specifically, the agent moves to a new location within the space, and disconnects from the current network then re-establish a network based on agents layout by using a cyclic timeout event with an interval of 2 yrs. As dynamic network, the fraction of \term{CS} and \term{CEU}s among its connected agents are modified with the change of the \term{SN}, therefore, influence the effect of \term{SN} on \term{SI} and \term{ECigUI}. 

\begin{figure}
\centering
\includegraphics[scale=0.28]{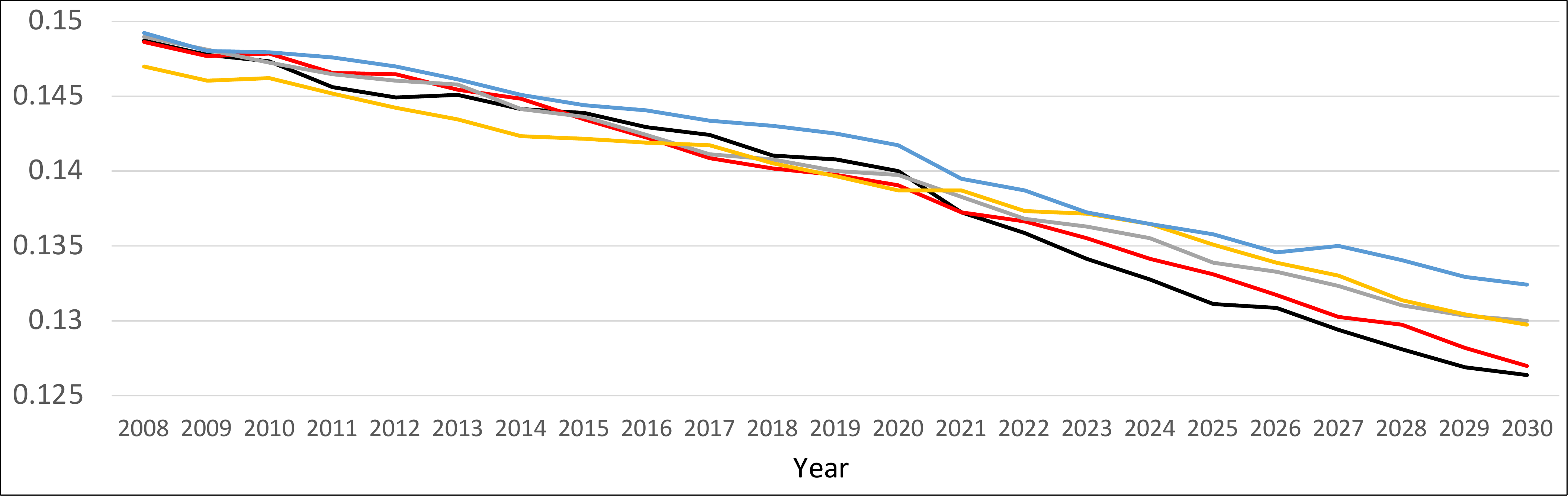}
\captionof{figure}{SA of rate of \term{ECigUC} on \term{PCS}. 
Black, red, grey, yellow, and blue line represent a successively larger rate of \term{ECigUC} of 0.2, 0.4, 0.6, 0.8, and 1.0, respectively.}  
\label{fig8}

\centering
\includegraphics[scale=0.28]{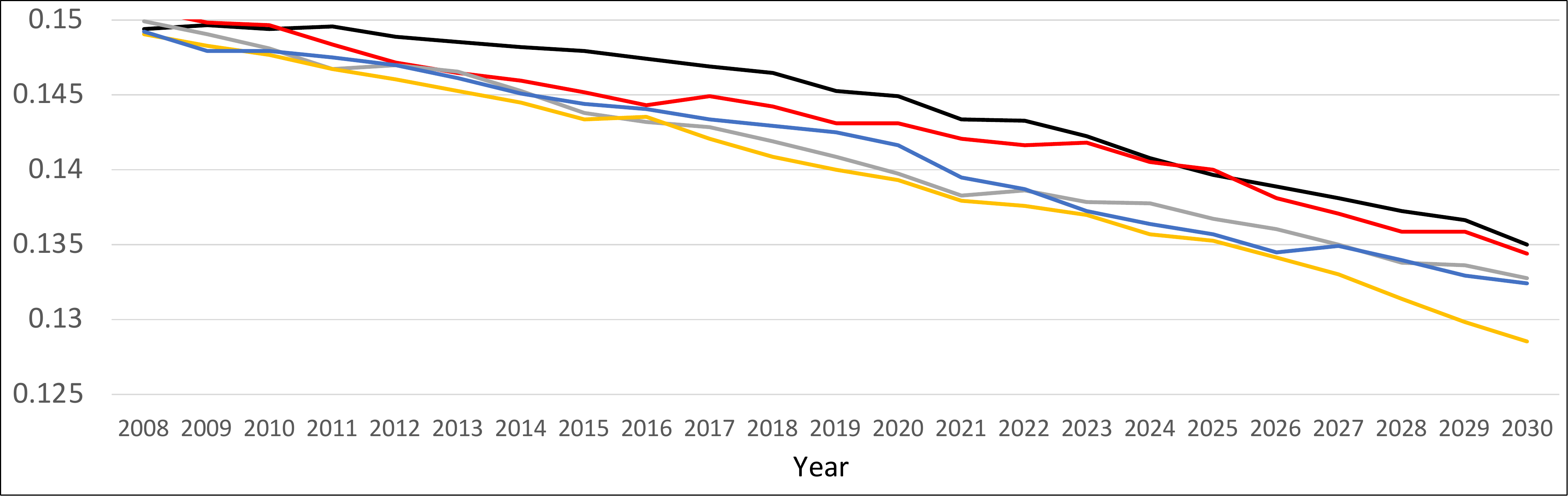}
\captionof{figure}
{SA of rate of \term{ECigUR} on \term{PCS}. Black, red, grey, yellow, blue line represent a successively larger rate of \term{ECigUR} of 0.2, 0.4, 0.6, 0.8, 1.0, respectively.}  
\label{fig9}

\centering
\includegraphics[scale=0.21]{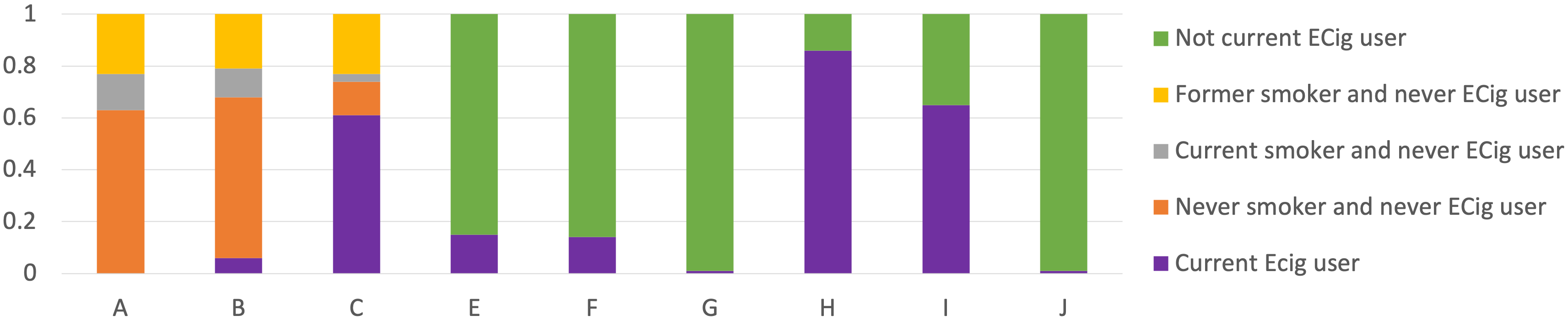}
\captionof{figure}{Panels A, B and C depict the population breakdown by smoking category in \term{Scn1}, \term{Scn2}, and \term{Scn3}, respectively. Panels E, F, G, H, I and J illustrate fraction of \term{CEU} among \term{CS}, among \term{FS} and among \term{NS} in \term{Scn2} and those fractions in \term{Scn3}, respectively.}
\label{fig5}
\end{figure}

The effect of \term{SN} was modeled using multipliers ($m\of{net}$), and applied them to the baseline $r\of{si}$ and $r\of{ECig}$, respectively. The overall rate of \term{SI} and \term{ECigUI} of a particular agent were increased by $m\of{net}$, relative to the rates in the baseline scenario. Without a specific mathematical model to quantify the effects of connected neighbors of a particular agent, the model employed a sigmoid function to describe the progression of the influence from the connected neighbors, which increases small at the beginning then accelerates fast and reaches the plateau. 
We, therefore, assumed $m\of{net}$ follows a sigmoid function (Equation \ref{eq3}), where $f$ is the fraction of \term{CS} (or, correspondingly, \term{CEU}s) among its connected agents if $m\of{net}$ is used to calculate the rate of \term{SI} or \term{ECigUI}, respectively, and $f_0$, $\alpha$ and $\gamma$ in Equation \ref{eq3} are 0.25, 2.0 and 1.0, respectively. Similarly, in Equation \ref{eq3}, $r\of{average}$ represents the average rate of \term{SI} or rate of \term{ECigUI} of population for the calculation of rate of \term{SI} or \term{ECigUI} of this agent, respectively. The $r\of{average}$s are re-calculated every year based on the smoking and ECig use status of the population at the beginning of each year. 

\begin{equation} \label{eq3}
    m\of{net} = \frac{\alpha + e^{-\gamma \times (f - f\of0)}} {(1 + e^{-\gamma \times (f - f\of0)}) \times r\of{average}}
\end{equation}

\module{Model Calibration}
$e\of{si}$, $e\of{sc}$, $e\of{sr}$ and $e\of{ECig}$ were calibrated to match the estimated \term{PCS}, prevalence of former smoker (\term{PFS}) and \term{PCEU} generated by the rates ($r\of{si}$, $r\of{sc}$, $r\of{sr}$ and $r\of{ECig}$) in the baseline model, against historical data of 2013-2017 from CTADS \cite{CTADS}. The calibrated result of $e\of{si}$, $e\of{sc}$, $e\of{sr}$ and $e\of{ECig}$ is 1.088, 2.435, 1.51 and 7.898, respectively.




\module{Model Scenarios} 
We examine here simulated population-level smoking behaviour change and ECig use under following three scenarios: smoking behavior in scenario one (\term{Scn1}) which is in absence of ECig use and the \term{SN}, smoking behavior in scenario two (\term{Scn2}) which is under the use of ECig, and smoking behavior in scenario three (\term{Scn3}) which \term{SN} exists and supports the \term{SI} and \term{ECigUI} (\term{Scn3}). 
The outputs from these scenarios examined the difference in prevalence and incidence of smoking arising from considering ECigs as well as \term{SN} both separately and in combination. The simulation of \term{Scn1} and \term{Scn2} were run for 100 realizations, and simulation under \term{Scn3} were run for 40 realizations with respect to the considerably large computation of \term{SN} in AnyLogic, with random seeds making each simulation run unique, then the means of the outputs of all runs were calculated for the comparison. Furthermore, to examine the statistical significance between the results from \term{Scn1} and \term{Scn2}, we performed a Mann-Whitney-U test on the per-realization output (\term{PCS}), from the two scenarios.

\module{Sensitivity Analysis}
To assess the sensitivity of model parameters on model outputs, we performed sensitivity analysis (SA) on the parameters such as the rate of \term{ECigUC} and the rate of \term{ECigUR}. 
The message transitions for \term{ECigUC} and \term{ECigUR} were replaced by the rate transitions. 
In the SA of the rate of \term{ECigUC}, the model assumed the rate of \term{ECigUR} is 1.0, and the range of the rate of \term{ECigUC} was 0.2 to 1.0 with a step of 0.2 for each iteration. Similarly, for the SA of the rate of \term{ECigUR}, the model assumed the rate of \term{ECigUC} is 1.0 and the rate of \term{ECigUR} had the same range and step with the rate of \term{ECigUC} in its SA experiment. The SA experiments examined the potential change of \term{PCS} resulting from changes in the value of the rate of \term{ECigUC} and the rate of \term{ECigUR}.

\section{Results}

\module{Comparison between \term{Scn1} and \term{Scn2}}
Mean, median and standard deviation of the results for \term{PCS}, generated by the model realizations in \term{Scn1}, are 0.1438, 0.1440 and 0.0037, respectively, and those from the model realizations in \term{Scn2} are 0.1369, 0.1374 and 0.0074, respectively. The results of a two-sided Mann-Whitney-U test for the results of two scenarios, $p < 2.2e^{-16}$, demonstrates that the distributions in the results of two scenarios differed significantly.



The message transitions in ECig use statechart were disabled in \term{Scn1} and \term{Scn2}, therefore, in the stacked column chart showing the breakdown by smoking category (Figure \ref{fig5}A, B and C), the agents were divided into four categories: \term{CEU} regardless of their smoking status, \term{NS}\aswellas\term{NEU}, \term{CS}\aswellas\term{NEU}, and \term{FS}\aswellas\term{NEU}, respectively. 
The portion of \term{FS} and \term{NEU} (23\%) in \term{Scn1} is slightly higher than that (21\%) in \term{Scn2}, due to a large portion (6\%) of \term{CEU}, as shown in Figure \ref{fig5} A and B.  This reflects the fact that the \term{FS} in \term{Scn1} is located within the \term{FS}\aswellas\term{NEU} category, whereas in \term{Scn2} some of those individuals are located within the \term{CEU} category.

\module{Comparison between \term{Scn2} and \term{Scn3}}
In \term{Scn3}, at the end of simulation, the maximum and minimum degree centrality of a given agent is 2 and 1, respectively. With the presence of \term{SN} (in \term{Scn3}), as shown in Figure \ref{fig5} C, the fraction of \term{CEU} in the population increased dramatically -- rising from 6\% in \term{Scn2} to 61\% in \term{Scn3}. With the exposure to ECig use from connected individuals or neighbors, people tend to initiate ECig use. The increased portion of \term{CEU} are mostly from the agents who were \term{NS}\aswellas\term{NEU}. 
In \term{Scn2}, the fraction of \term{CEU} among \term{CS} and that among \term{FS} are similar, with value of 15\% and 14\% in Figure \ref{fig5} E and \ref{fig5} F, respectively, which are considerably larger than fraction of \term{CEU} among \term{NS}, as shown in Figure \ref{fig5}G. In \term{Scn3}, the \term{SN} significantly increased the fraction of \term{CEU} among \term{CS} and \term{FS}, with the value of 86\% and 65\%, respectively in Figure \ref{fig5}H and Figure \ref{fig5}I, respectively, while the fraction of \term{CEU} among \term{NS} does not show obvious increase due to \term{SN}, compared with that of \term{Scn2}.


\module{Sensitivity Analysis}
Results from the SA on the rate of \term{ECigUC} and the rate of \term{ECigUR} suggests that the \term{PCEU} and prevalence of former ECig user can substantially change the \term{PCS}, as shown in Figure \ref{fig8} and \ref{fig9}. Figure \ref{fig8} demonstrates that when \term{ECR} is increased from 0.2 to 1.0 -- holding invariant the value of \term{ERR} -- \term{PCS} are gradually increased, and \term{PCEU} decreases.
The results in Figure \ref{fig8} suggest that although incidence of \term{SI} is reduced by the lower \term{PCEU}, the decreased rate of \term{SC} and elevation in \term{SR} due to the decreased \term{PCEU} compensates for the decrease in the rate of \term{SI}. 
Similarly, the change in the rate of \term{ECigUR} also influences \term{PCEU}.  Holding constant the rate of \term{ECigUC}, an increase in the rate of \term{ECigUR} generally increases the \term{PCEU}, but lowers the \term{PCS}, with a possible exception at the lowest levels of the rate of \term{ECigUR}. Results in Figure \ref{fig9}, the line from the rate of \term{ECigUR} of 0.2 having the lowest \term{PCEU}, reflect that agents were more likely to remain as \term{CS}.

\section{Discussion}
From the results in the three scenarios, the model demonstrates that ECig use and \term{SN} encourage agents to uptake ECig, therefore, shape population-level smoking behavior. Although the use of ECig increases the rate of \term{SI}, the combined effect of the increase in the rate of \term{SC} and the decrease in the rate of \term{SR} results a considerably large decline in \term{PCS} and increase in \term{PFS}. The results of SA further shows the \term{PCS} is sensitive to the ECig use behavior change. The outputs of the model largely depend on the feedback between smoking and ECig use, and interactions between agents. 
First, we assumed the rate of \term{ECigUI} of \term{CS}, \term{FS} and \term{NS} are in a declining order, specifically, the \term{CS} has highest rate of \term{ECigUI} compared with other smoking category. Second, if an individual is \term{CS}, being a \term{CEU} increases the probability of quitting smoking and staying in \term{FS} state, which means they have a relatively higher probability of using ECig as a \term{SC} tool. We assumed the ECig use helps greatly in \term{SI} for \term{NS}. Given the model results, fraction of \term{CEU} among \term{NS} is considerably lower, compared with \term{CS} and \term{FS}. Accordingly, as a combined result of the rate of \term{SI}, the rate of \term{SC} and rate of \term{SR}, the \term{PCS} is decreased due to ECig use. Furthermore, we assumed gender effect as divisors in the rate of \term{ECigUI}. Thus, the model behaves a relatively stronger influence from ECig use on smoking behavior. The effect of \term{SN} is modeled as a multiplier to the rate of \term{ECigUI}, which generates more \term{CEU} during simulation.

Despite fine resolution of the model, there are some limitations. First, the model is highly sensitive with the use of ECig, however, the model has no good assumption on the rate of \term{ECigUC} and the rate of \term{ECigUR}. Second, at this resolution, the model cannot capture the smoking episodes, dynamics of nicotine metabolism, allowing model to analyze whether ECig use helps in relieving nicotine cravings at fine-grained level as \term{SC} tool. Finally, the model assumes the effect of \term{SN} in a relatively simple way.

Although with some limitations, the model outcomes can provide some straightforward understanding of the complex feedback between smoking and ECig use at individual level, then allow us to analyze population-level smoking behaviour. Additionally, the model is also a useful tool for examining how \term{SN} influences smoking and ECig use, particularly among adolescents.

%
%

%
%
%
\bibliographystyle{splncs04}
\bibliography{sbp2019_ecig}

\end{document}